\documentclass[prd,reprint,letterpaper,superscriptaddress,amsfonts,amsmath,amssymb,nofootinbib]{revtex4-2}

\usepackage{braket}

\usepackage{graphicx}
\usepackage{color}
\usepackage[normalem]{ulem}
\usepackage{MnSymbol}
\usepackage{enumerate}
\usepackage{bbold}
\usepackage{subfigure}
\usepackage{subfiles}
\usepackage{units}
\usepackage[usenames,dvipsnames]{xcolor}
\usepackage[colorlinks=true,linkcolor=MidnightBlue,citecolor=blue,filecolor=green,urlcolor=Violet]{hyperref}
\usepackage{colortbl}
\usepackage[dvipsnames]{xcolor}
\usepackage{paralist}
\usepackage{psfrag}
\usepackage{setspace,tabularx,fancyhdr}
\usepackage{mathtools}
\usepackage[top=2cm,bottom=2cm,left=2cm,right=2cm]{geometry}
\usepackage{enumitem}
\setlist[itemize]{leftmargin=*}
\usepackage{wrapfig}
\usepackage{type1cm}
\usepackage{yfonts}
\usepackage{float}

\def\beq{\begin{equation}}
\def\eeq{\end{equation}}
\def\bsp{\begin{split}}
\def\esp{\end{split}}
\def\bea{\begin{eqnarray}}
\def\eea{\end{eqnarray}}

\definecolor{myyellow}{rgb}{0.94, 0.86, 0.51}

\definecolor{mygreen}{rgb}{0.2, 0.8, 0.2}

\definecolor{mypink}{rgb}{0.99, 0, 0.99}

\definecolor{mypurple}{rgb}{0.75, 0, 0.75}

\definecolor{cadmiumorange}{rgb}{0.93, 0.53, 0.18}

\newcommand{\IGNORE}[1]{}

\graphicspath{{Images/}}

\begin{document}

\title{Constraining Black Hole Horizon Properties Through Long-Duration Gravitational Wave Observations}

\author{Ikram Hamoudy}\email{ikramhamoudy25@gmail.com}
\affiliation{Department of Physics, Bar-Ilan University, Ramat Gan 5290002, Israel}

\author{Julian Westerweck}
\affiliation{Institute for Gravitational Wave Astronomy and School of Physics and Astronomy, University of Birmingham, Edgbaston, Birmingham B15 2TT, United Kingdom}
\affiliation{Max-Planck-Institut f\"ur Gravitationsphysik (Albert-Einstein-Institut),
Callinstra{\ss}e 38, D-30167  Hannover, Germany}
\affiliation{Leibniz Universit\"at Hannover, D-30167 Hannover, Germany}

\author{Ofek Birnholtz}
\affiliation{Department of Physics, Bar-Ilan University, Ramat Gan 5290002, Israel}

\begin{abstract}
We perform a long-duration Bayesian analysis of gravitational-wave data to constrain the near-horizon geometry of black holes formed in binary mergers. Deviations from the Kerr geometry are parameterized by replacing the horizon’s absorbing boundary with a reflective surface at a fractional distance $\epsilon$. This leads to the emission of long-lived monochromatic quasinormal modes, which can be probed with extended integration times. Building on previous work that set a bound of $\log_{10} \epsilon$ = $\text{-}24$ for GW150914, we reproduce and validate those results, and extend the analysis to additional events from LIGO–Virgo–KAGRA observing runs. By combining posterior samples from multiple detections, we construct a joint posterior yielding a tightened $90\%$ upper bound of $\log_{10}\epsilon < \text{-}38.64$, demonstrating the statistical power of population-level inference through cumulative evidence. Finally, analyzing the newly observed high signal-to-noise ratio event GW250114 \cite{LVK_GW250114} from the O4b run, we obtain the most stringent single-event constraint to date, $\log_{10}\epsilon < \text{-}29.58$ (90\% credible region). Our findings provide the strongest observational support to date for the Kerr geometry as the correct description of post-merger black holes instead of a class of horizonless compact objects, with no detectable horizon-scale deviations.

\end{abstract}
\maketitle

\section{Introduction}\label{sec:Introduction}
The advent of gravitational wave (GW) astronomy has opened an unprecedented window into the universe, allowing us to probe the nature of compact objects and the dynamics of extreme astrophysical events. The detection of GW150914 \cite{LigoCollab}, the first observation of GWs from the merger of two black holes by the LIGO–Virgo collaboration, provided strong evidence for the existence of black holes described by the Kerr solution, as predicted by General Relativity (GR)  \cite{kerr1963}. However, it also raised intriguing questions about the fundamental nature of these objects, particularly the structure of their horizons.

In GR, a black hole’s event horizon acts as a perfect absorber. The remnant of a binary merger settles into equilibrium by emitting gravitational radiation at specific frequencies known as quasinormal modes (QNMs), whose spectrum is uniquely determined by the black hole’s mass and spin. Alternative scenarios propose horizonless ultracompact objects in which the horizon is replaced by a reflective surface at a small fractional distance $\epsilon$ outside the Kerr radius. In this case, gravitational perturbations can be temporarily trapped and re-emitted in additional modes, as long-lived, nearly monochromatic GWs. For the parameters of GW150914, such modes would appear at $\sim 210$ Hz, well within the LIGO band, with lifetimes ranging from tens to thousands of seconds \cite{UCO}.

Several searches for black hole echoes and near-horizon modifications have been carried out  \cite{events2021, yes_echoes, no_echoes, no_echoes2}, with different results: while some analyses reported hints of non-Kerr structures, others found no statistically significant evidence. Recent studies \cite{Reconstructed_paper,Mondal:2025uxm} introduced a long-duration Bayesian analysis to test for such signals for GW150914, placing an upper bound of $\log_{10} \epsilon$ = $\text{-}23.5$ ($90\%$ credible region), and later incorporating further detections. This corresponds to ruling out reflective surfaces down to subatomic scales \cite{Medved2019}. The additional modes in this framework are damped sinusoids with frequency, damping time, and amplitude fixed by the geometry and the black hole’s parameters.

The motivation for this research is to validate and extend these findings. By analyzing long post-merger data segments from multiple GW detections, including high signal-to-noise ratio (SNR) events from the first part of the LIGO-Virgo-KAGRA (LVK) collaboration’s fourth observing run (O4a) \cite{gwtc4_intro,gwtc4_results}, we aim to impose tighter constraints on $\epsilon$. 
Similarly to those previous works, we assume reflective boundary conditions at the near-horizon boundary, and note it can be motivated by the extremely weak interactions of GWs with matter.
We note that we explicitly exclude material bodies extending far outside the would-be-horizon, such as neutron stars \cite{Maggio:2017ivp}, as such bodies, and objects without a clear photon sphere (non-ClePhOs \cite{UCO}), can be probed by other tests.
We also note that while perfect reflectivity can induce ergoregion instability \cite{real+im}, an absorption of $\lesssim0.5\%$ should be enough to quench it for objects of the spins we consider here; and such low absorption also means relatively little loss of signal amplitude \cite{Testa:2018bzd}.
We defer an exact analysis of small non-zero reflectivity to future work.

In this work, we reproduce the analysis of GW150914, extend it to additional events, and explore how factors such as SNR and time-window selection influence the strength of the inferred bounds. To further strengthen the conclusions, we combine posterior samples from multiple events, constructing a joint posterior that yields a substantially improved upper bound on $\epsilon$, $\log _{10} \epsilon$ = -38.64, when including all the most promising events from observing runs O1 through O4a and one from O4b. Excluding the O4 events yields a slightly weaker bound of 
$\log _{10} \epsilon$ = -36.2, demonstrating the statistical power of population-level inference. 

With the publication of the first GW event from the second part of the LVK's fourth observing run (O4b) \cite{LVK_GW250114,Abac2025}, and anticipating new high-SNR detections \cite{GW80},  even more precise tests of black hole structure are becoming possible \cite{Abac2025}.
Notably, the event GW250114, with a network SNR of approximately 80 \cite{LVK_GW250114, Abac2025, GW80}, provides the strongest single-event constraint to date, yielding $\log_{10}\epsilon < \text{-}29.58$. This represents an improvement of almost six orders of magnitude over the original GW150914 result, pushing the sensitivity of near horizon tests to unprecedentedly low scales.

\section{Background \& Literature Review}\label{sec:Background}
According to GR, astrophysical black holes are expected to possess an event horizon that acts as a perfect absorber. The exterior vacuum geometry of a stationary, rotating black hole is described by the Kerr solution \cite{kerr1963}. However, reconciling GR with quantum mechanics raises the information loss problem: Hawking radiation appears to carry no information about the black hole’s interior \cite{paradox,loss_info}. One class of proposed resolutions involves modifying the classical picture by introducing a near-horizon surface at a small fractional distance $\epsilon$ from the Kerr horizon \cite{resolved_paradox}, replacing the perfectly absorbing boundary with a partially or fully reflecting one.

We parametrize deviations from the Kerr geometry through a shifted surface at radius
\bea
R &=& r_+  (1+\epsilon),
\label{eq: shifted R} \\
\epsilon &=& \frac{\Delta R}{r_+},
\label{eq: epsilon}
\eea
where $R$ is the near horizon surface, which is positioned at a relative distance $\epsilon$ above $r_+$, the location of the would-be horizon. Measuring $\epsilon$ directly tests whether the object is a classical black hole or a horizonless ultracompact object (UCO) \cite{UCO55}.

The background spacetime is described in Boyer–Lindquist coordinates \cite{Boyer} (we use $G=c=1$):
\bea
ds^2 = &-& \left( 1 - \frac{2Mr}{\Sigma} \right) dt^2 
 - \frac{4Mr}{\Sigma} a \sin^2 \theta \, d\phi dt 
 + \frac{\Sigma}{\Delta} dr^2 + \Sigma d\theta^2 \nonumber\\
 &+& \left( (r^2 + a^2) \sin^2 \theta 
 + \frac{2Mr}{\Sigma} a^2 \sin^4 \theta \right) d\phi^2,
\label{eq: Boyer-Lindquist}
\eea
where $M$ is the black hole mass, $a$ is the spin parameter, and
\bea
\Sigma &=& r^2 + a^2 \cos^2 \theta, \nonumber\\
\Delta &=& r^2 + a^2 - 2Mr = (r - r_+)(r - r_-), \nonumber\\ 
r_\pm &=& M \pm \sqrt{M^2 - a^2}  .
\eea

When two black holes merge, the perturbed remnant relaxes by emitting QNMs — damped oscillations whose spectrum depends only on the mass and spin of the black hole \cite{QNM2018, mode2021}. Standard QNM analyses probe the geometry near the light ring \cite{LR}, but are insensitive to microscopic near-horizon modifications. To test this regime, we impose modified boundary conditions for perturbations at 
$r=R$, corresponding to a reflecting surface.

\begin{figure} [H] 
    \begin{centering}    \includegraphics[width=0.45\textwidth]{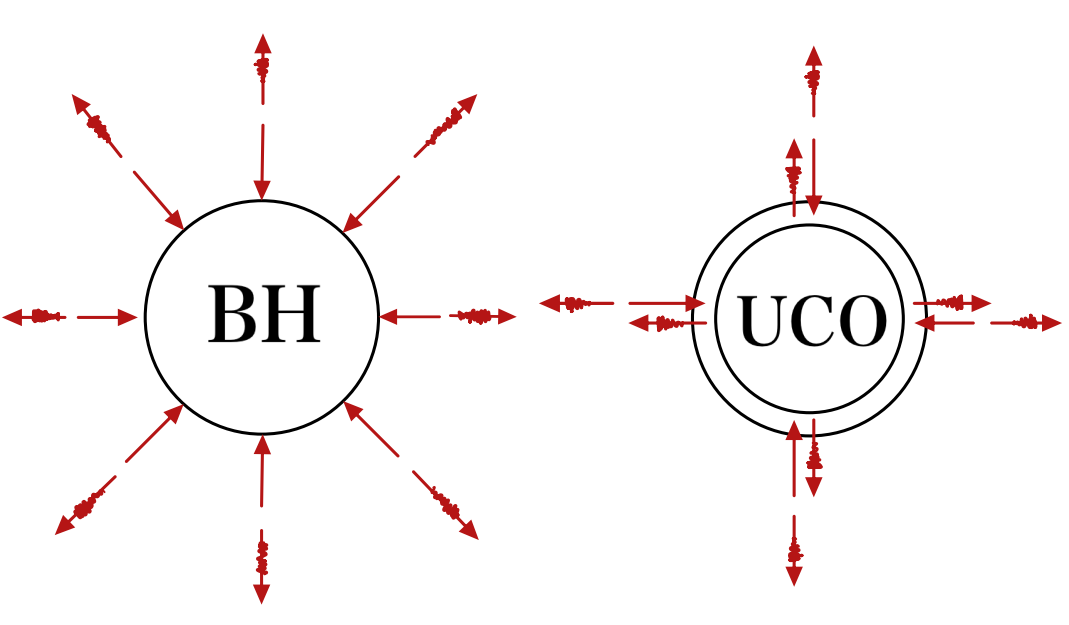}
    \caption{The figure compares the behavior of GWs in black holes and UCOs. In BHs, infalling GWs are fully absorbed at the horizon, leaving no reflection. In contrast, for UCOs the GWs are weakly interacting, and reemerge after a delay due to partial reflection from a surface just outside the would-be horizon. This delayed re-emission is caused by the high gravitational redshift near the surface. This property is used during the post-merger ringdown to examine the presence or absence of a horizon.}
    \label{fig: UCOvsBH}
    \end{centering}    
\end{figure}

This behavior can be described by an effective model in which a reflective surface is located a small but finite distance $\epsilon > 0$  outside the horizon. The extreme gravitational redshift near this surface introduces a significant time delay, but finite 
$\epsilon $ ensures that waves re-emerge within a finite time as observed from infinity. This parameterization provides a framework to model horizonless ultracompact objects and quantify deviations from the classical Kerr picture \cite{Medved2019}.

Imposing reflective boundary conditions in the Kerr geometry leads to the appearance of long-lived, nearly monochromatic QNMs that emerge after a time delay and dominate the late-time gravitational-wave emission. These modes are physically related to, but distinct from, the so-called black hole echoes. Both arise from near-horizon reflections and produce extended post-merger signals. Unlike phenomenological echo models \cite{Abedi}, which approximate the reflection pattern with a sequence of repeating signals, the present model yields an analytically derived single damped sinusoid determined entirely by the reflective boundary conditions and black hole parameters. The extra free parameters in echo searches reflect modeling uncertainties rather than fundamental physical freedom.

Several searches for black hole echoes have been conducted using data from GW detectors like LIGO and Virgo. Some analyses have reported tentative evidence for echoes, suggesting possible near-horizon structures \cite{yes_echoes}, while others have found no statistically significant evidence \cite{no_echoes, no_echoes2}.

The recently observed GW250114 event has become the loudest binary black hole merger detected to date, with a network SNR of approximately 80 \cite{LVK_GW250114}.  
It features nearly equal component masses with a total mass of $65.8^{+1.1}_{-1.2}$ M$_\odot$ and a mass ratio $q \gtrsim 0.91$ \cite{Abac2025}.  
The low component spins yield a particularly clean ringdown signal, enabling accurate mode identification.  
Follow-up analyses \cite{Lu2025} have identified multiple QNMs and a potential direct-wave component oscillating near twice the horizon’s angular frequency ($2\Omega_H$), consistent with theoretical expectations for frame dragging near the horizon.  
These characteristics make GW250114 an exceptional probe for tests of the Kerr geometry and horizon-scale structure, providing the strongest single-event constraint yet on $\log_{10}\epsilon$.

Perturbations of the Kerr geometry satisfy the Teukolsky equation \cite{Teukolsky}, which in tortoise coordinates 
$dr_\ast/dr = (r^2 + a^2)/\Delta$ takes the form of a potential scattering problem:
\beq
 \frac{d_s^2\Psi_{lm}}{dr_\ast^2} - V(r,\omega)_s\Psi_{lm} = 0.
\label{eq: perturbations}
\eeq
This equation describes potential scattering in flat space, and the effective potential $V(r,\omega)$ appears in \cite{Potential}. The indices $l$ and $m$ label the angular harmonic components of the perturbation, while the spin $s = \pm2$ for gravitational perturbations. Its solutions behave approximately as follows:
\bea
 \Psi  \sim e^{i\omega r_\ast} \quad &,&\quad r_\ast \rightarrow \infty,
\label{eq: sol1} \\
 \Psi  \sim e^{-i\omega r_\ast} + \mathcal{R} e^{i\omega r_\ast}  \quad  &,&\quad r_\ast \rightarrow r_\ast (R),
\label{eq: sol2}
\eea
where $\mathcal{R}$  represents the reflection coefficient of the surface, and the complex frequency $\omega$ is expressed as $\omega = \omega_R + i\omega_I$. It is worth mentioning, according to the transformation above, if $\epsilon\to 0$, then $r_*(R)\to -\infty$.

For this study, we adopt $\mathcal{R}=1$, corresponding to a perfectly reflecting surface used as an idealized limit to maximize potential near-horizon effects. While the Einstein equivalence principle implies that GWs interact only weakly with matter \cite{Partial_absorption, fuzzball}, this assumption models cases where waves pass through the interior without significant interaction, leading to an effective reflection, and provides a benchmark for testing deviations from the Kerr geometry. The opposite limit, $\mathcal{R}=0$, corresponds to a fully absorbing Kerr black hole.

\begin{figure}[H] 
    \begin{centering}
   \includegraphics[width=0.5\textwidth]{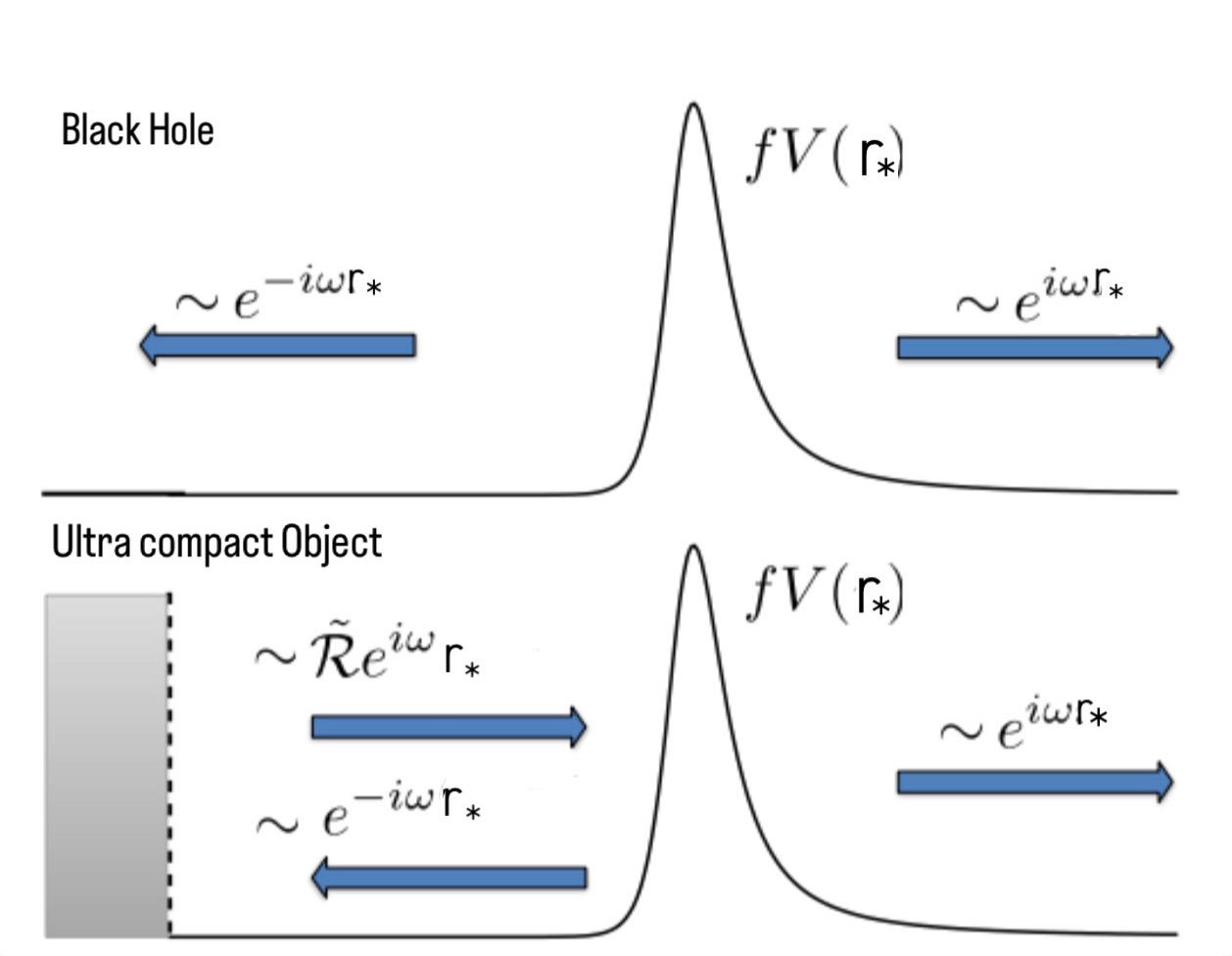}
    \caption{Top panel: The boundary conditions that govern wave propagation in black hole spacetime. Bottom panel: The reflecting conditions for waves outside an UCO. Figure taken from \cite{boundary}.}
 
    \label{fig: wave emission}
    \end{centering}    
\end{figure}

This modification leads to additional long-lived, nearly monochromatic QNMs. For $\epsilon \ll 1$, the real and imaginary parts of the dominant mode frequency are given by \cite{real+im, weak}:
\bea
  \omega_R  &\simeq& m\Omega \pm \frac{\pi}{2|r_\ast^0|} (\nu+1),
\label{eq: real} \\
  \omega_I &\simeq& \frac{2M(\omega_R-m\Omega)r_+}{225 |r_\ast^0| 
 (r_+-r_-)}[\omega_R(r_+-r_-)]^5,
\label{eq: imaginary}
\eea

where $|r_\ast^0| \sim M(1 + ( 1 - \chi ^2)^{-1/2})|\ln \epsilon|$ is the cavity proper length. The angular velocity of the horizon ($\Omega$) is related to the spin parameter $\Omega = (a/M)/2r_+ = \chi/2r_+ $ and $\chi = a/M$ is the dimensionless spin parameter ($a$ which initially has units of mass, is divided by mass). 
The overtone number of the additional long-lived quasinormal modes is $\nu$, not to be confused with the standard QNMs' overtone index $n$.
These modes have frequencies near the black hole’s rotational frequency and damping times scaling as $\tau  \sim r_+ |\ln \epsilon|^2$.

For GW150914 parameters, this corresponds to frequencies $\sim 210$Hz and lifetimes from tens to thousands of seconds — within LIGO’s sensitive band.

The resulting GW signal is a damped sinusoid,
\begin{equation} \label{eq: waveform}
       (h_+ + ih_x)(t) = S_{lm}Ae^{-(t - t_0)/\tau}  e^{i(2\pi f (t-t_0)+ \phi )} \Theta(t-t_0),
\end{equation}

 fully determined by the black hole parameters and $\epsilon$, unlike typical pulsed-echo models. Detecting or constraining this signal allows us to probe the near-horizon structure with unprecedented precision. 

\section{Waveform Model}
\subsection{Signal Description}

To model the late-time GW emission, we use an analytical signal model that captures the damped oscillations of QNMs excited after the merger. Referring to Eq.~\ref{eq: waveform}, we focus on the dominant spherical mode \( l = m = 2 \), approximating spheroidal harmonics by spin-weighted spherical harmonics \cite{Harmonics}, and the dominant overtone $\nu = 1$. For \( \epsilon \ll 1 \) the real part of the frequency is mainly determined by the angular velocity of the object. Therefore, we specialize Eq.~\ref{eq: real} to the dominant gravitational perturbation, and by using the near-horizon cavity length and the relation between the Kerr horizon angular velocity and the spin parameter, Eq.~\ref{eq: real} simplifies to the form in Eq.~\ref{eq: freq}. In this expression, the real part of the mode frequency explicitly depends on the black hole spin \( \chi \) and the fractional near-horizon offset \( \epsilon \). Thus, the dominant frequency can be written as
\begin{equation} \label{eq: freq}
     M\omega_R = \frac{\chi}{1 + \sqrt{1 - \chi^2}} 
     + \frac{\pi \sqrt{1 - \chi^2}}{|\ln \epsilon|(1 + \sqrt{1 - \chi^2})} \, .
\end{equation}

The amplitude of the additional long-lived modes is determined by the total energy flux carried away as GWs \cite{energy}. This energy consists of the prompt post-merger GW emission and the rotational energy of the compact object ($\Delta E = E_{\text{initial}} + E_{\text{rotational}} $). The initial energy of the additional modes can be estimated using the GR prediction that the inward GW energy flux toward the compact surface is approximately equal to the prompt outward emission \cite{Medved2019}. The rotational energy is evaluated using a simplified non-relativistic approximation, $E_{\text{rotational}}$ = $\frac{1}{2}$ $E_{\text{initial}} \Omega ^2 r_+^2$. This, combined with the relation between $\Omega$ and the dimensionless spin parameter $\chi$, gives the total available energy for the additional modes,
\beq
    \Delta E = E_{\rm initial}\left(1 + \frac{\chi^2}{8}\right),
    \label{eq: energy}
\eeq
where \( E_{\rm initial} \) represents the energy flux immediately after the merger, estimated to be comparable to the energy radiated during the prompt emission \cite{Medved2019}. For GW150914, this corresponds to roughly three solar masses.

The dominant contribution to the signal traverses the effective potential barrier of the Kerr spacetime, whose peak can be estimated using a first-order WKB approximation \cite{WKB} as
\[
  V_{\max} \approx (M\omega_{\rm QNM})^2, \qquad M\omega_{\rm QNM} \sim 0.5.
\]
Here, $\omega_{\text{QNM}}$ is the fundamental Kerr frequency of the quasi-normal mode.
After the ingoing ringdown radiation pulse is reflected in the near-horizon interior region, it encounters the potential barrier and is partially transmitted and reflected.
Its high-frequency components escape more efficiently, leading to initial pulsed echoes through subsequent traversals of the the cavity formed by interior reflector and potential barrier, while the low-frequency components are trapped, forming a standing wave in the cavity.
These low-frequency modes are the quasi-normal modes of the geometry modified by the inner reflection and slowly leak out, leading at late times to the long-lived quasi-monochromatic signal we target here~\cite{Cardoso:2016oxy}.

The GW energy flux at infinity is given by
\begin{equation} \label{eq: derivative}
    \frac{dE_{\rm GW}}{dt} 
    = \frac{D_L^2}{32\pi} \int \langle \dot{h}_{\mu \nu} \dot{h}^{\mu \nu} \rangle d\Omega ,
\end{equation}
where \( D_L \) is the luminosity distance and angle brackets denote short-wavelength averaging. For a nearly monochromatic signal with \( \omega_R \simeq 2\Omega \), this integral simplifies to 
\(\dot{E}_{\rm GW} \approx \frac{1}{4} D_L^2 \langle |\dot{h}|^2 \rangle\). Combining this with the waveform model in Eq.~(\ref{eq: waveform}) for \( \omega_R \tau \gg 1 \), the amplitude becomes
\beq
     A = \frac{4}{\omega_R D_L} \left( \frac{\Delta E}{\tau} \right)^{1/2} .
\label{eq: amplitude}
\eeq
We can see the parameters of this equation specified in Eqs. (\ref{eq: real}), (\ref{eq: imaginary}) and (\ref{eq: energy}), with the explicit expression for $\omega_R$ given in Eq.~(\ref{eq: freq}).

The damping time \(\tau\) for these modes is
\begin{equation} \label{eq: tau}
     \tau = \frac{225M}{32\pi} 
     \left( \frac{1 + \sqrt{1 - \chi^2}}{\sqrt{1 - \chi^2}} \right)^6
     \frac{|\ln \epsilon|^7}{(\chi |\ln \epsilon| + \pi \sqrt{1 - \chi^2})^5}.
\end{equation}

Following \cite{Reconstructed_paper}, the start time of the analysis is set to \( t_0 = 2\,\mathrm{s} \) after the merger to avoid contamination from standard ringdown modes while retaining most of the signal energy. Although the reflection at the light-ring barrier is weak, a fraction of the wave can remain trapped near the compact object. For configurations with small $\epsilon$, these residual perturbations decay slowly (long damping time) and can dominate the late-time signal, motivating the choice \( t_0 = 2\,\mathrm{s} \) to focus on the long-lived component while excluding the prompt ringdown.

\subsection{Assumptions}

The model is built under the following assumptions:
\begin{itemize}
  \item The final object is treated as a horizonless ultracompact object, replacing the absorbing boundary at the horizon with a perfectly reflecting surface. This allows deviations from the Kerr geometry to be encoded in long-lived monochromatic QNMs.
  \item The surface is modeled as perfectly reflecting, motivated by the Einstein equivalence principle: any partial reflection (\(0 < \mathcal{R} < 1\)) would imply violations of the principle.
  \item The system is assumed planar, with GWs emitted and reflected in the same plane.
  \item All infalling and extracted energy is re-emitted as the described long-lived mode, ignoring potential losses through early post-merger emission such as pulsed echoes or higher mode contributions. This would enter as a scaling factor into available energy and thus expected amplitude and procured bounds on deviations.
\end{itemize}

\section{Data Analysis Framework}

To constrain deviations from the Kerr geometry, we employ a Bayesian inference framework applied to long-duration post-merger GW data. The analysis uses publicly available strain data from the LVK detectors and selected high–SNR events from the O1–O4a observing runs and GW250114 from O4b. The goal is to probe potential near-horizon modifications by searching for long-lived, monochromatic QNMs predicted for reflective boundary conditions.

\subsection{Data Selection} 

We analyze strain data from the Gravitational-Wave Open Science Center (GWOSC) \cite{GWOSC1,GWOSC2,GWOSC3,GWOSC}, focusing on binary black hole mergers with high expected post-merger SNR. In particular, GW150914 is used to reproduce and validate previous constraints \cite{Reconstructed_paper}. We then extend the analysis to an additional 18 public events from O1–O3 and three high-SNR detections from the O4a run, GW230814\_230901, GW231226\_101520 and GW231206\_233901.

For the exceptionally loud event GW230814 (SNR = 42) \cite{GW230814}, we additionally examine successive 128-second segments of post-merger data (0–128 s, 128–256 s, etc.). This extended analysis leverages the high SNR to search for potential late-time signals to probe for delayed modes while controlling noise contamination at late times. 

Finally, we analyze the recently reported event GW250114 (SNR $\sim$ 80), the loudest binary black hole detection to date. Owing to its exceptional signal strength, this event provides a unique sensitive test of horizon-scale deviations and enables a substantially tighter constraint on $\log_{10} \epsilon$.

\subsection{Bayesian Inference and Implementation} 

We perform parameter estimation using the PyCBC Inference toolkit \cite{Method2,Histogram}, which provides Bayesian posterior sampling through a parallel-tempered Markov-chain Monte Carlo algorithm \cite{Method3}. The likelihood and priors follow the standard formulation:
  \begin{equation} \label{eq:  posterior}
       p(\Vec{\theta}|d,h) = \frac{p(d|\Vec{\theta},h) p(\Vec{\theta}|h)}{p(d|h)}
\end{equation}
where $d$ is the data, $\vec{\theta}$ is the model parameters, $h$ is the signal model, in this case for the UCO's long-lived mode \cite{Method1}. Inspiral–merger–ringdown (IMR) posteriors are used to set priors on mass, spin, inclination, and distance \cite{PE}, and calculate the expected energy emitted in the long-lived mode.
We choose a uniform distribution of $\log_{10} \epsilon \in [-45,-2]$ as the only additionally prescribed prior.

Finally, posterior samples of 
$\log _{10} \epsilon $ are extracted and processed to compute $50\%$ and $90\%$ credible intervals. We also construct combined posteriors by multiplying posteriors from independent events, thereby tightening bounds through cumulative evidence.

\subsection{Analysis Pipeline}
The analysis pipeline proceeds as follows:
\begin{enumerate}
\item Acquiring and preprocessing strain from GWOSC.

\item Estimating IMR parameters using PyCBC to obtain posteriors on mass, spin, and radiated energy.

\item Post-merger Bayesian analysis on successive time windows using the reflective-boundary waveform model to infer $\epsilon $.

\item Population-level combination of posteriors to obtain joint constraints on $\epsilon $.

\end{enumerate}

This pipeline is applied uniformly across all events to ensure consistency and comparability.

\section{Combining GW Event Bounds} \label{sec:combining}
To improve constraints on near-horizon deviations beyond what is achievable with individual events, we combine posterior distributions on the parameter $\log _{10} \epsilon$ obtained from multiple GW detections. Assuming that the measured physical quantity is shared between individual events, each provides an independent measurement, allowing us to statistically tighten the bounds by aggregating information across the population \cite{static}.

Under the assumption that (i) the same physical model applies to all events, (ii) the prior on 
$\log _{10} \epsilon$ is identical, and (iii) the measurements are statistically independent, the combined posterior is constructed by multiplying the individual posteriors for each event \cite{Method4}:
\begin{equation} \label{eq:com_posterior}
p(\theta | d_1, d_2, \ldots, d_n) \propto \prod_{i=1}^{n} p(\theta | d_i),
\end{equation}
As before, $d$ is the data, ${\theta}$ is the model parameters ($\log _{10} \epsilon$), and $d_i$ is the data from the $i$-th GW event.

Each individual posterior is represented using kernel density estimation (KDE), which provides a smooth, non-parametric approximation suitable for accurate multiplication. This procedure follows standard Bayesian methodology for combining independent measurements and has been widely adopted in GW tests of general relativity.

The main advantage of this approach is that it fully exploits the information content of all available events, resulting in a significantly narrower credible interval on $\log_{10} \epsilon$ compared to any single-event analysis. In particular, combining high-SNR events leads to substantially tighter upper bounds, thereby improving sensitivity to potential near-horizon deviations.

\section{Results} \label{sec:results}

\begin{table*}[t]
\centering
\begin{tabular}{l c c c c}
\hline\hline
Run & Event name ~~~ & ~~~ Simulated signal SNR ~~~ & ~~~ IMR SNR ~~~& Bound ($\log_{10}\epsilon$) \\
\hline\hline
\multicolumn{5}{l}{\textbf{Part A : O1, O2, and O3 events}} \\
\hline
O1 & GW150914\_095045 & 5.10 & 26.0 & -24.15 \\
\hline
O2 & GW170104\_101158 & 3.21 & 13.8 & -15.61 \\
   & GW170823\_131358 & 3.07 & 12.2 & -14.11 \\
   & GW170814\_103043 & 3.04 & 17.7 & -18.06 \\
\hline
O3a & GW190521\_074359 & 4.42 & 25.9 & -22.23 \\
    & GW190727\_060333 & 3.99 & 11.7 & -14.15 \\
    & GW190828\_063405 & 3.15 & 16.5 & -16.66 \\
    & GW190630\_185205 & 3.14 & 16.4 & -16.10 \\
    & GW190503\_185404 & 3.10 & 12.2 & -16.15 \\
    & GW190915\_235702 & 2.99 & 13.1 & -16.73 \\
    & GW190910\_112807 & 2.55 & 14.5 & -15.60 \\
\hline
O3b & GW200129\_065458 & 6.67 & 26.8 & -25.08 \\
    & GW200112\_155838 & 3.50 & 19.8 & -17.77 \\
    & GW200128\_022011 & 3.76 & 10.6 & -13.33 \\
    & GW200311\_115853 & 3.65 & 17.8 & -18.71 \\
    & GW191222\_033537 & 3.14 & 12.5 & -13.44 \\
    & GW200219\_094415 & 2.95 & 10.7 & -13.09 \\
    & GW200224\_222234 & 2.61 & 20.0 & -19.70 \\
\hline    
\multicolumn{5}{l}{\textbf{Part B : O4a events}} \\
\hline
O4a & GW231206\_233901 & 4.87 & 21.9 & -23.74 \\
    & GW230814\_230901 & 4.14 & 43.0 & -20.31 \\
    & GW230927\_153832 & 4.36 & 20.3 & -21.72 \\
    & GW231226\_101520 & 3.95 & 34.7 & -19.34 \\
\hline
\multicolumn{5}{l}{\textbf{Part C : O4b event}} \\
\hline
O4b & GW250114\_082203 & 9.91 & 80.0 & -29.58 \\
\hline\hline
\end{tabular}
\caption{Summary of constraints on the horizon distance parameter $\log_{10}\epsilon$ for GW events across observing runs O1--O4b. Lower values of $\log_{10}\epsilon$ indicate smaller deviations from a perfectly absorbing Kerr horizon, with stronger constraints obtained for events with higher SNR. The table additionally lists the SNR for a simulated post-merger signal of fiducial parameters, and for the inspiral-merger-ringdown (IMR) signal.}
\label{tab:summary}
\end{table*}

\subsection{Individual Event Analyses} \label{subsec:individual}
We first apply the analysis pipeline to 18 public gravitational-wave events from the GWOSC database, spanning observing runs O1–O3. For each event, we perform Bayesian parameter estimation on post-merger data segments to constrain the near-horizon deviation parameter $\log_{10} \epsilon$. Table~(\ref{tab:summary}), Part A, provides a summary of the constraints derived from the O1–O3b events, reporting the corresponding merger SNR values, expected SNR estimates from simulated long-duration signals, and the bounds found from analyzing the data.

\begin{figure}
\centering
\includegraphics[width=0.45\textwidth]{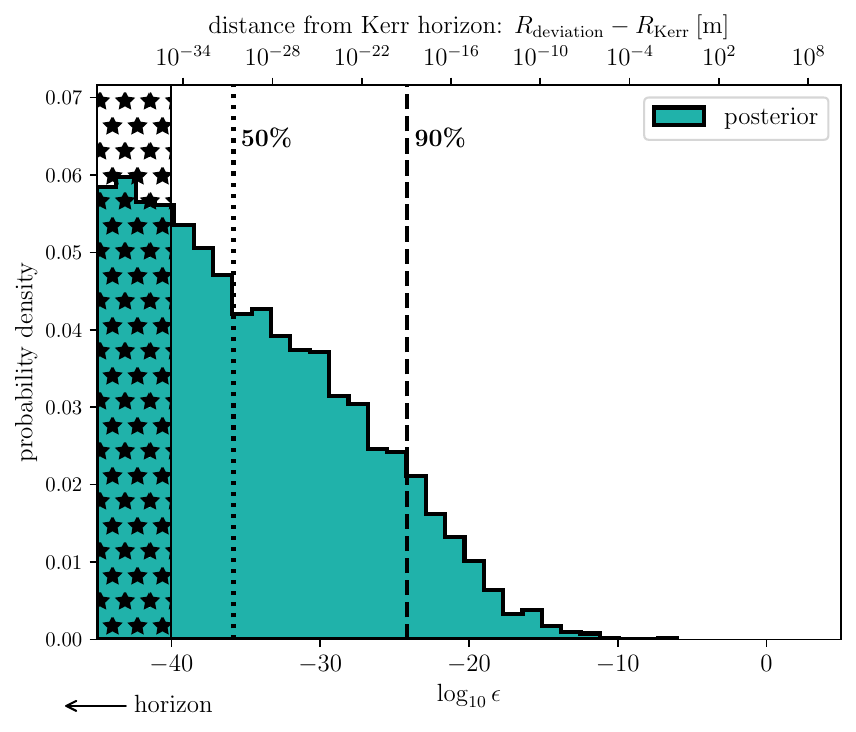}
s\caption{Posterior distribution of $\log _{10} \epsilon$ for GW150914. Shaded regions indicate probability density; dashed (dotted) lines mark $90\%$ ($50\%$) credible intervals. The leftmost hatched region corresponds to Planckian scales (the Planck length from the classical horizon $\sim 10^{-35}$m shown in the top axis). The posterior exhibits the same qualitative behavior observed across all analyzed events — a peak near the Kerr horizon and a rapid decay toward larger $\epsilon$ values.}
\label{fig:plots}
\end{figure}

Figure~\ref{fig:plots} presents the posterior distributions for GW150914 (the same analysis for all analyzed events).

Across the sample, the $50\%$ credible intervals are consistently concentrated near the classical Kerr horizon, while the $90\%$ intervals extend further, indicating that the data favor at most extremely small deviations from the Kerr geometry, but with event-dependent uncertainties.

The best $90\%$ credible region bound from O1 - O3 observing runs, $\log _{10} \epsilon <$ \text{-}25.08, is obtained for GW$200129$.

In particular, GW200129 exhibits a moderately positive effective inspiral spin ($\chi_{\mathrm{eff}} = 0.15$) and inclination ($\iota = 0.80$), both of which contribute to stronger excitation and improved visibility of the dominant quasinormal mode content following merger. These properties increase the amplitude and coherence of the ringdown signal observed in the detectors, facilitating tighter constraints on post-merger deviations from the Kerr geometry, and explain why GW$200129\_065458$ outperforms other events with comparable or even larger merger SNRs (see Table~\ref{tab:summary}, Part B).

\begin{figure}
\centering
\includegraphics[width=0.45\textwidth]{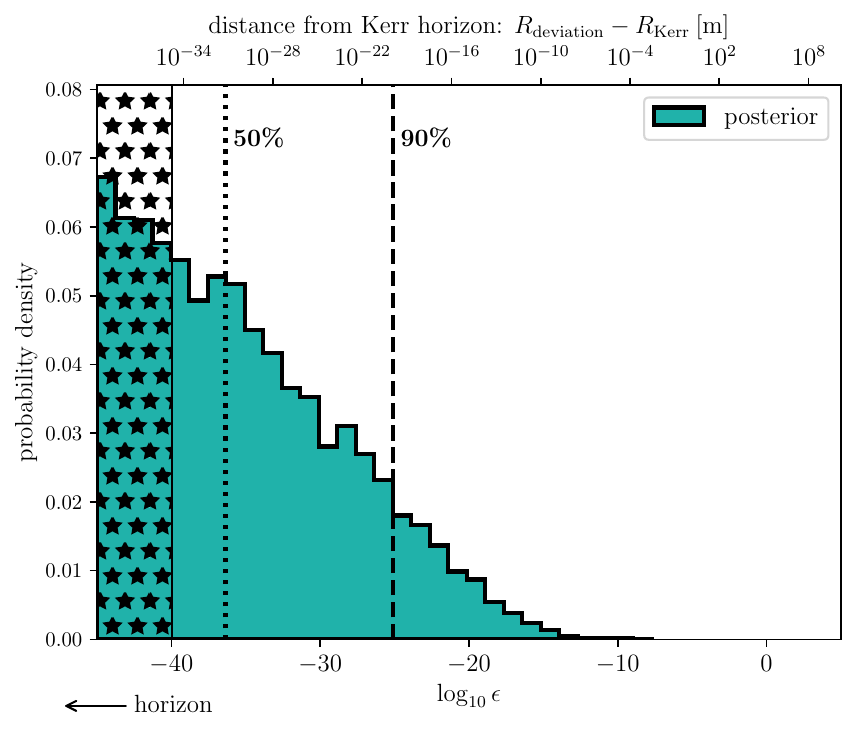}
\caption{Posterior distribution of $\log _{10} \epsilon$ for GW$200129$. Shaded regions indicate probability density; dashed (dotted) lines mark $90\%$ ($50\%$) credible intervals. The leftmost hatched region corresponds to Planckian scales (the Planck length from the classical horizon $\sim 10^{-35}$m shown in the top axis). The posterior exhibits the same qualitative behavior as previously observed for GW150914.}
\label{fig:GW200129posterior}
\end{figure}

The resulting posterior distribution for $\log _{10} \epsilon $, displayed in Figure~\ref{fig:GW200129posterior}, shows a sharply localized peak near the Kerr limit, with probability density rapidly diminishing toward larger deviations. This behavior confirms that the data provide a strong preference for horizon-scale structure consistent with a classical Kerr black hole.

\begin{figure*}
\centering
\includegraphics[width=0.92\textwidth]{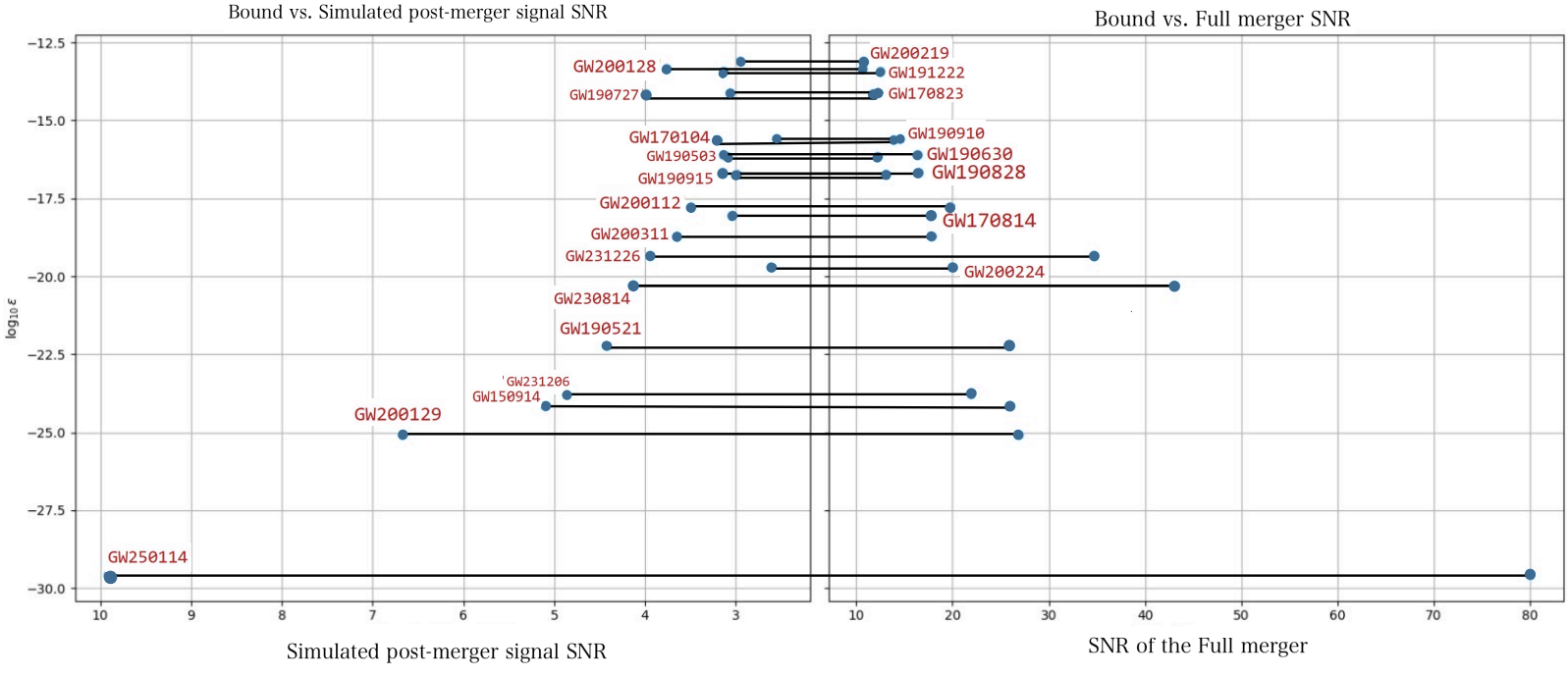}
\caption{Scatter plots comparing the obtained upper bounds on $\epsilon$ to the SNR of each event. \textbf{Left:} Bound vs.\ simulated long-lived mode SNR. Note the inverted horizontal axis. \textbf{Right:} Bound vs.\ IMR SNR. For each event, two linked markers are shown — one for the simulated-mode SNR and one for the IMR SNR — and each pair is labeled with the event name. Events with higher SNRs provide tighter bounds, moving toward smaller values of $\epsilon$. The left panel shows a wider scatter at low simulated-mode SNRs, indicating that noise dominates the bound determination in those cases.}
\label{fig:Scatter}
\end{figure*}

The relationship between the inferred bounds and the signal strength is shown in Figure~\ref{fig:Scatter}.
The dependence on the SNR for the signal from the black hole merger is approximately linear, reflecting the statistical scaling of parameter uncertainty with SNR. In contrast, the dependence on the SNR for the simulated long-duration signal shows increased scatter at low SNR due to noise-dominated behavior.

\subsection{High-SNR Events from O4a}
\begin{figure}
\centering
\includegraphics[width=0.47\textwidth]{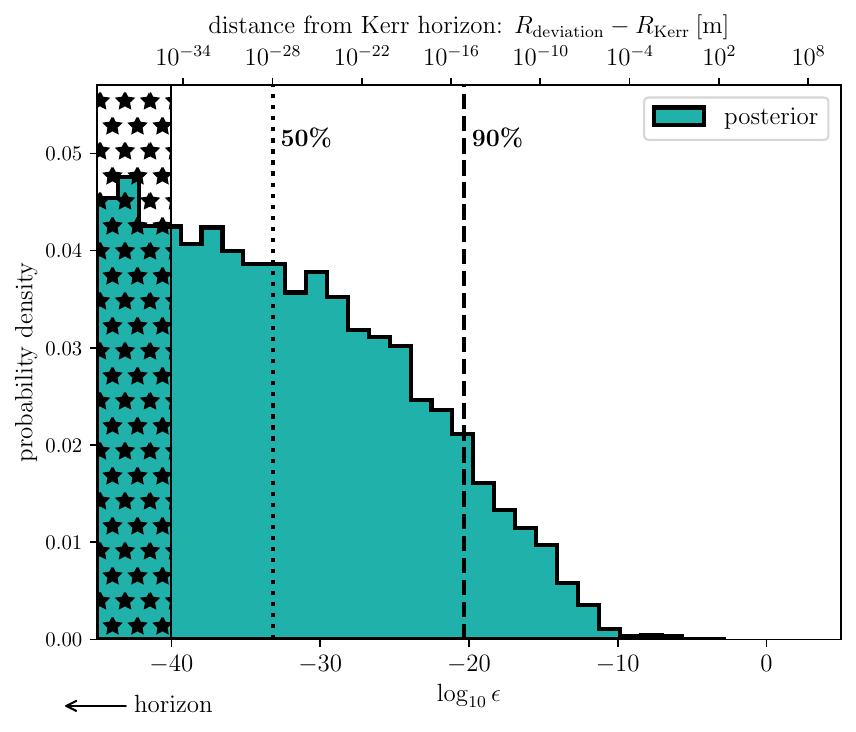}
\caption{Posterior distribution of $\log _{10} \epsilon$ for GW230814. The vertical dashed (dotted) lines indicate the $90\%$ ($50\%$) credible intervals.}
\label{fig:O4plots}
\end{figure}

We extended the analysis to four high-SNR events from the LVK O4 observing run:
GW$230814\_230901$ (shown in Figure~\ref{fig:O4plots}), GW$231206\_233901$,
GW$231226\_101520$, GW$250114_082203$ (Figure~\ref{fig:GW250114}).

Table \ref{tab:summary} Part B reports the constraints obtained for these events. The bounds remain broadly consistent with previous runs.

To probe potential late-time deviations, we analyzed the highest-SNR O4a event (GW$230814\_230901$) across multiple 128-second post-merger segments.  
The obtained bounds were:
\[
\begin{aligned}
-128{-}0~\text{s} &\Rightarrow \log_{10}\epsilon = -20.30,\\
0{-}128~\text{s} &\Rightarrow \log_{10}\epsilon = -20.31,\\
128{-}256~\text{s} &\Rightarrow \log_{10}\epsilon = -19.12,\\
256{-}384~\text{s} &\Rightarrow \log_{10}\epsilon = -20.63,\\
384{-}512~\text{s} &\Rightarrow \log_{10}\epsilon = -19.96,\\
512{-}640~\text{s} &\Rightarrow \log_{10}\epsilon = -19.60.
\end{aligned}
\]

These bounds remain broadly consistent, with no significant temporal evolution or systematic trend.
This stability suggests that the remnant behaves in accordance with the Kerr solution and that no persistent echo-like features are present.  
The slightly weaker bound in the second window ($-19.12$) likely reflects statistical fluctuations or transient increases in detector noise.  

The pre-merger control window ($-128$ – $0$~s) yielded a similar bound ($-20.30$), confirming that these fluctuations are consistent with background noise rather than physical signals. Hence, there is \textbf{no evidence for long-lived echoes} or other post-merger deviations in the analyzed data.

\subsection{Combined Bounds Across Events}
\begin{figure}
\centering
\includegraphics[width=0.5\textwidth]{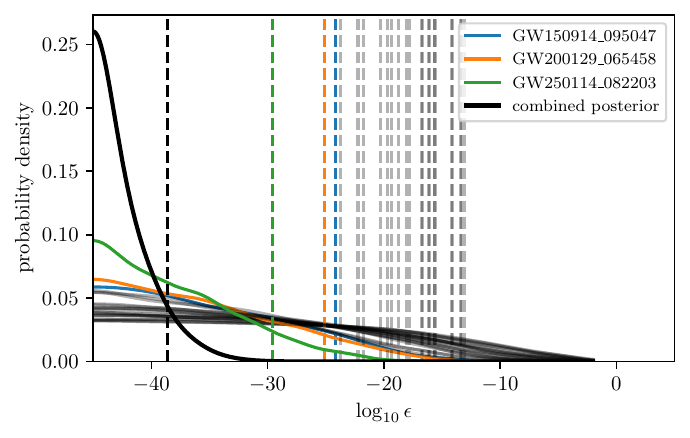}
\caption{Combined posterior for $\log_{10}\epsilon$ (black line), compared to the single-event posteriors (light grey, most constraining posteriors colored). One-sided $90\%$ credible upper bounds are shown by dashed lines, with the combined posterior yielding $\log_{10} \epsilon = -38.64$.}
\label{fig:Combining}
\end{figure}

To leverage available information, we combine posterior distributions across events from O1 - O4 using the procedure described in Sec.~\ref{sec:combining}. The resulting combined posterior, shown in Figure~\ref{fig:Combining}, is sharply peaked at lower values of $\log_{10}\epsilon$, yielding a $90\%$ upper bound of (\text{-}38.64). 

\begin{figure}
\centering
\includegraphics[width=0.5\textwidth]{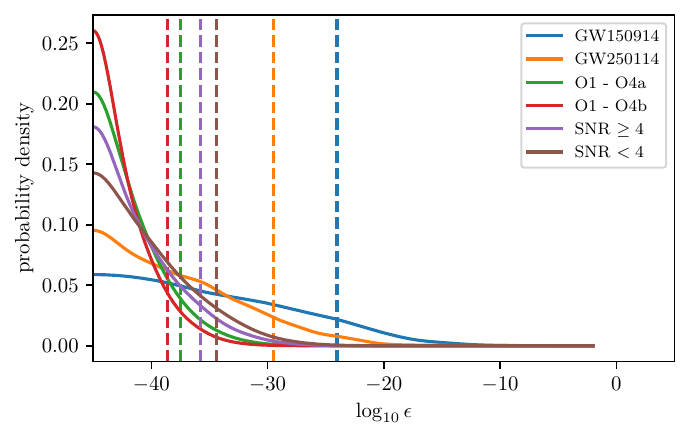}
\caption{Combined posteriors including various selections of events, including selected by the predicted SNR in the long-lived mode (see Table~\ref{tab:summary}). Solid lines show posterior distributions, dashed lines the corresponding $90\%$ upper bounds. This illustrates both the effectiveness of including high-SNR events and larger numbers of low-SNR events.}
\label{fig:Posterior_combinations}
\end{figure}

This represents an improvement of many orders of magnitude compared to the best single-event bound (\text{-}29.58) of the combined events, and tighter than the original GW150914 analysis (\text{-}24).
The effect of including different selections of events in a hierarchical posterior is shown in Figure~\ref{fig:Posterior_combinations}, illustrating the value of recent high-SNR detections and larger populations of low-SNR detections in improving our bounds.

\subsection{High-SNR Event from O4b: GW250114}

\begin{figure}
    \centering
    \includegraphics[width=0.5\textwidth]{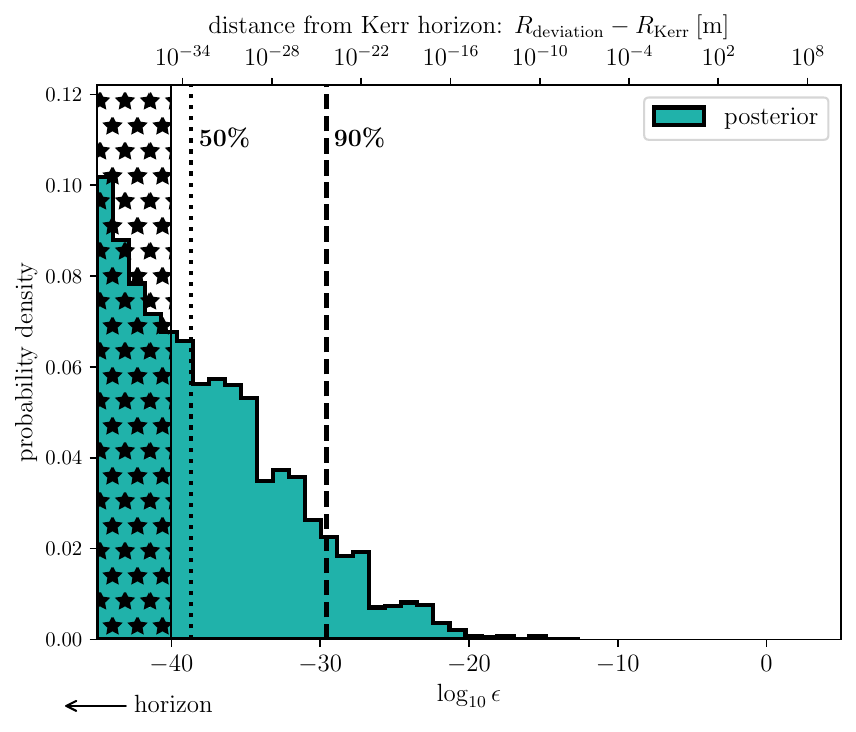}
    \caption{Posterior distribution of $\log_{10}\epsilon$ for the GW250114 event (O4b). The $90\%$ ($50\%$) credible intervals are shown by the dashed (dotted) vertical lines, with the $90\%$ upper bound at $\log_{10}\epsilon = -29.58$.}
    \label{fig:GW250114}
\end{figure}

The newly reported \textbf{GW250114} event, detected during the O4b run with a network SNR of approximately 80, provides the strongest single-event constraint to date, shown in Figure~\ref{fig:GW250114}. Applying the same post-merger analysis yields a $90\%$ credible upper bound of
\[
\log_{10}\epsilon = -29.58,
\] 
This result surpasses the previous best single-event constraint (from GW200129, $\log_{10}\epsilon = -25.08$) by over four orders of magnitude.

The posterior distribution peaks sharply near the horizon, indicating no detectable deviation from the Kerr geometry, and places the modification to the horizon near Planck-scale distances — approaching the Planck length $\sim 10^{-35}$m from the classical horizon. The posterior exhibits the same qualitative behavior as observed for previously analyzed detections.

\section{Discussion and Conclusions}
This work constrained potential deviations from the Kerr black hole geometry by analyzing long-duration post-merger gravitational-wave signals and evaluating the horizon-shift parameter $\log_{10}\epsilon$. Using Bayesian inference across multiple events, separately and combined, we tested for horizon-scale modifications such as echoes and assessed the consistency of observed remnants with the Kerr solution.

Our key findings are:
\begin{itemize}
    \item The bounds scale inversely with the SNR, as expected from statistical principles. 
    \item Multi-event combination (O1 - O4) strengthens constraints, yielding a joint bound of $\log_{10}\epsilon < -38.64$. 
    \item The highest-SNR O4b event, GW250114, provides a new best single-event constraint of $\log_{10}\epsilon < -29.58$, tightening previous limits by almost six orders of magnitude.
\end{itemize}

The methodology was validated by reproducing previous results for GW$150914$. Extending the analysis to 18 additional events from the O1–O3 runs yielded consistent bounds that scale inversely with SNR, confirming that higher-SNR events provide tighter constraints. The strongest individual limit, $\log_{10}\epsilon < -25.08$, was obtained for GW$200129\_065458$, likely due to its favorable spin and inclination, which enhance late-time ringdown visibility.

For the loudest O4a event, GW$230814\_230901$ (SNR = 42), we examined successive 128-second post-merger windows up to 640 seconds. The bounds remained stable ($\log_{10}\epsilon \sim -20$), showing no evidence of echoes or time-dependent deviations, and confirming that the remnant behaves as a classical Kerr black hole.

Combining posteriors from O1 - O4 events produced a significantly tighter bound,
\[
\log_{10}\epsilon < -38.64,
\]
This result underscores the statistical power of population-level inference and supports the Kerr geometry as a robust description of astrophysical black holes.

These results support the robustness of the Kerr metric and show no evidence for long-lived post-merger echoes or exotic compact object behavior. The GW250114 result places the horizon/reflection distance close to Planckian scales, approaching $\sim 10^{-35}$m from the classical horizon. The posterior exhibits the same qualitative behavior as those obtained for previous detections, providing the strongest observational evidence to date for the classical black hole paradigm.

Future analyses will focus on combining O4b events, including GW250114, within a unified Bayesian framework to derive a population-level constraint. Given the scaling of sensitivity with SNR, integration time and event count, and assuming the arrival of a few very high-SNR post-merger signals combined with improved detector performance, these analyses will tighten the bounds beyond those obtained in this work, probing unprecedented regimes of near-horizon physics.

\section{Acknowledgments}
I.~H.\ has been supported by the Israel Science Fund (ISF) grant No.\ 1698/22. and by the US-Israel Binational Science Foundation (BSF) grant No.\ 2024816, as well as by a scholarship program for Arab minority research students provided by the Israeli Council for Higher Education's Planning \& Budgeting Committee.
J.~W.\ acknowledges support through the UK STFC grant ST/W000946/1.

This research has made use of data or software obtained from the Gravitational Wave Open Science Center (gwosc.org), a service of the LIGO Scientific Collaboration, the Virgo Collaboration. This material is based upon work supported by NSF's LIGO Laboratory which is a major facility fully funded by the National Science Foundation, as well as the Science and Technology Facilities Council (STFC) of the United Kingdom, the Max-Planck-Society (MPS), and the State of Niedersachsen/Germany for support of the construction of Advanced LIGO and construction and operation of the GEO600 detector. Additional support for Advanced LIGO was provided by the Australian Research Council. Virgo is funded, through the European Gravitational Observatory (EGO), by the French Centre National de Recherche Scientifique (CNRS), the Italian Istituto Nazionale di Fisica Nucleare (INFN) and the Dutch Nikhef, with contributions by institutions from Belgium, Germany, Greece, Hungary, Ireland, Japan, Monaco, Poland, Portugal, Spain.

\end{document}